\documentstyle{article}
\begin{document}
\title{Structure of Spinning Particle \\
Suggested by Gravity, Supergravity \\
and Low Energy String Theory}
\author{ A. Burinskii \\
Gravity Research Group, NSI Russian Academy of Sciences,\\
 B. Tulskaya 52,  Moscow 113191, Russia, e-mail: grg@ibrae.ac.ru\thanks
{Talk at the International Workshop Spin'99, Prague, 5-11 September, 1999}}
\date{October, 1999}
\maketitle
\begin{abstract}
  The structure of spinning particle suggested
by the rotating  Kerr-Newman ( black hole ) solution, super-Kerr-Newman
solution and the Kerr-Sen solution to low energy string theory is
considered.
Main peculiarities of the Kerr spinning particle are discussed: a
vortex of twisting principal null congruence, singular ring
and the Kerr source representing a rotating relativistic disk of the
Compton size. A few stringy structures can be found in the real and complex
Kerr geometry.
    Low-energy string theory predicts the existence of a heterotic
string placed on the sharp boundary of this disk.  The obtained recently
supergeneralization of the Kerr-Newman solution suggests the existence of
extra axial singular line and fermionic traveling waves concentrating near
these singularities.
\par
      We discuss briefly a possibility of experimental test of these
 predictions.
\end{abstract}
\section{Introduction}
 The Kerr solution is well known as  a  field  of  the
rotating black hole. However, for the case of  a  large  angular
momentum $L$, $\mid a\mid  = L/m \geq  m$,  all the horizons of the Kerr
metric are absent, and the  naked  ring-like  singularity is appeared.
This naked singularity has many unpleasant manifestations and must be
hidden  inside a rotating disk-like source. The Kerr solution  with $\mid
a\mid  \gg  m$   displays some remarkable features  indicating  a relation
to the structure  of   the   spinning   elementary particles.
\par
In the 1969
Carter  \cite{car} observed, that if three parameters of the Kerr - Newman
solution are adopted to be ($\hbar $=c=1 ) $\quad e^{2}\approx  1/137,\quad
m \approx 10^{-22},\quad a \approx  10^{22},\quad L=ma=1/2,$
then one obtains  a  model  for  the four  parameters  of  the electron:
charge, mass, spin and  magnetic moment, and the giromagnetic ratio is
automatically the same as that of the Dirac electron. Israel \cite{is}
has introduced a disk-like source for the Kerr field, and it was shown by
Hamity \cite{ham} that this source represents a rigid relativistic rotator.
 A model of "microgeon" with Kerr metric was suggested \cite{bur}  and
an analogy  of this model  with string models \cite{ibur}. Then a model of
the Kerr-Newman source in the form of oblate spheroid was suggested
\cite{lop}. It was shown that material of the source  must  have very
exotic properties: null energy density and negative  pressure.
An attempt to explain these properties
on the basis  of the  volume  Casimir effect was given in work \cite{blet}.
 The electromagnetic properties of the material  are close to those of a
superconductor \cite{blet,lop}, that allows to consider singular ring of the
Kerr source as  a closed vortex string like the Nielsen-Olesen  and Witten
superconducting strings.
Since 1992 black holes have paid attention of string theory.
In 1992 the Kerr solution was generalized by
Sen to low energy string theory \cite{sen}.
It was shown that black holes can be considered as fundamental
string states, and the point of view has appeared that some of black holes
can be treated as elementary particles \cite{part}.
The obtained recently super-Kerr-Newman solution \cite{superBH,skn}
represents a natural combination of the Kerr spinning particle and
superparticle models and predicts the existence of extra axial singularity
and fermionic traveling waves on the Kerr-Newman background.
\section{Kerr singular ring}
 The Kerr string-like singularity appears in the rotating BH solutions
instead of the point-like  singularity of the non-rotating BH.
The simple solution possessing the Kerr singular ring was
obtained by Appel in 1887 (!) \cite{app}. It can be considered
as a Newton or Coulomb analogue to the Kerr solution.
When the point-like source of the Coulomb solution
$f=1/\tilde r= 1/ \sqrt{ (x-x_o)^2+(y-y_o)^2 +(z-z_o)^2}$
is shifted to a complex point of space
$(x_o , y_o , z_o )\rightarrow (0,0,ia)$, the Kerr singular ring arises
on the real slice of space-time. The complex equation of singularity
$\tilde r=0$ represents a ring as an intersection of plane and sphere.
The complex radial distance $\tilde r$ can be expressed in the oblate
spheroidal coordinates $r$ and $\theta$:  $\tilde r = r+i a \cos \theta$.
The Kerr singular ring is a branch line of the space
on two sheets: "positive" one covered by $r \geq 0 $, and "negative" one,
an anti-world, covered by $ r \leq 0 $.
The sheets are connected by disk $ r = 0 $ spanned by singular ring.
The physical fields change signs and directions on the "negative "sheet.
Truncation of the negative sheet allows one to avoid the twosheetedness.
In this case the fields will acquire a shock crossing the disk, and some
 material sources have to be spread on the disk surface to satisfy
the field equations. The structure of electromagnetic field
near the disk suggests then that the "negative" sheet of space can be
considered as a mirror image of the real world in the rotating
superconducting mirror.
\par
The source of Kerr-Newman solution, like the Appel solution,
can be considered from complex point of view as a "particle" propagating
along a complex world-line $x^i (\tau)$
\cite{bkp,c-str} parametrized by complex time $\tau$.
  The objects described by the
complex world-lines occupy an intermediate position between  particle
and  string. Like the string
they form the two-dimensional surfaces or the world-sheets in the
space-time. It was shown that the complex Kerr source may be considered as
a complex hyperbolic string which requires an orbifold-like structure of
the world-sheet. It induces a related orbifold-like structure of the Kerr
geometry \cite{c-str} which is closely connected with the above mentioned
twosheetedness.
\section{Kerr congruence and disk-like source}
 Second remarkable peculiarity of the Kerr solution is the twisting
principal null congruence (PNC) which can be considered as a vortex
of null radiation. This vortex propagates via disk from negative sheet
of space onto positive one forming a caustic at singular ring.
 PNC plays fundamental role in the structure of the
Kerr geometry. The Kerr metric can be represented in the Kerr-Schild form
$g_{ik} = \eta _{ik} + 2 h k_{i} k_{k}, $
where $\eta $ is metric of an auxiliary Minkowski space and $h$ is a scalar
function. Vector field $ k_i(x) $ is null,
$ k_i k^i=0, $ and tangent to PNC.
The Kerr PNC is geodesic and shear free \cite{dks}. Congruences with such
properties are described by the Kerr theorem
 via complex function $Y(x)$ representing a projective spinor coordinate
$Y(x)= {\bar \Psi}^{\dot 2}/{\bar \Psi} ^{\dot 1}$.
The null vector field $k_i (x)$ can be expressed in spinor form
$k \sim \bar \Psi \sigma _i dx^i \Psi $.

The above complex representation of source allows one to obtain the
Kerr congruence by a retarded-time construction \cite{bkp,c-str}.
The complex light cone with the vertex at some point
$x_0$ of the complex world line
$(x_i - x_{0 i})(x^i -x_0^i) = 0$
can be split into two families of  null planes:  "left" and "right".
In spinor form this splitting can be described as
\begin{equation}
x_i = x_{0i} + \Psi \sigma _i \tilde \Psi ,\label{Psi}
\label{lk}\end{equation}
where "right" (or "left")  null planes  can be obtained
keeping  $\Psi$ constant and varying  $\tilde \Psi$, or keeping
$\tilde \Psi$ constant and varying  $ \Psi$.
The rays of the twisting Kerr congruence arise as {\it real slice} of the
"left" null planes of the complex light cones emanated from the complex world
line \cite{bkp,c-str}.
\par
Replacement of the negative sheet by a disk-like source at surface $r=0$
allows one to avoid twosheetedness of the Kerr space. However, there is still
a small region of causality violation on positive sheet of space.
By the L\"opez suggestion this region has to be also covered by source
\cite{lop}. The minimal value of $r$ covering this region is `classical
radius' $r_e=\frac{e^2}{2m}$. The resulting disk-like source has a thickness
of order $r_e$ and its degree of oblateness is $\alpha ^{-1} \approx 137$.
\section{Stringy suggestions}
 In 1974, in the frame of Einstein gravity the model of microgeon with the
Kerr-Newman metric was considered \cite{bur}, where singular ring was used
as a waveguide for wave excitations. It was recognized soon \cite{ibur} that
singular
ring represents in fact a string with traveling waves. Further, in dilaton
gravity, the string solutions with traveling waves have paid considerable
attention. The obtained by Sen generalization of the Kerr solution to low
energy string theory with axion and dilaton \cite{sen} was
analyzed in \cite{bsen}. It was shown that, in spite of the strong
deformation of metric by dilaton (leading to a change the type of
metric from type D to type I), the Kerr PNC survives in the
Kerr-Sen solution and retains the properties to be geodesic and shear
free. It means that the Kerr theorem and the above complex
representation are valid for the Kerr-Sen solution too.
It has also been obtained that the field of the Kerr-Sen solution near
the Kerr singular ring is similar to the field around a fundamental
heterotic string that suggested stringy interpretation of the Kerr
singular ring.
\section{Supergeneralization}
  Description of spinning particle based only on the bosonic
fields cannot be complete. On the other hand the fermionic models of
spinning particles and superparticles based on Grassmann coordinates
have paid considerable attention. In \cite{superBH,skn} a natural way to
combine
the Kerr spinning particle
and superparticle models was suggested leading to a non-trivial
super-Kerr-Newman black hole solution.
\par
The simplest consistent supergeneralization of Einstein gravity
represents an unification of the gravitational field $g_{ik}$,
 with a spin 3/2 Rarita-Schwinger field $\psi _i$ .
 There exists the problem of triviality of supergravity
solutions. Any exact solution of Einstein gravity is indeed a
trivial solution of supergravity field equations with a zero
field $\psi _i$. Starting from such a solution and
 using supertranslations, one can easily turn the gravity solution into a
form containing the spin-3/2 field $\psi_i$.
 However, since this spin-3/2 field can be gauged away by reverse
transformations such supersolutions have to be considered as {\it trivial}.
\par
The hint how to avoid this triviality problem was given by complex
representation of
the Kerr geometry. One notes that from complex
point of view the Schwarzschild and Kerr geometries are equivalent and
connected by a {\it trivial} complex shift.
The {\it non-trivial} twisting structure of the Kerr geometry arises as a
result of the {\it shifted real slice} regarding the source \cite{bkp,c-str}.
\par
Similarly, it is possible to turn a {\it trivial} super black hole
solution into a {\it non-trivial} if one finds an analogue to the {\it real
slice} in superspace.
\par
 The {\it trivial supershift} can be represented as a
replacement of the complex world line by a superworldline
$X^i_0(\tau)= x^i_0(\tau)-i \theta \sigma ^i \bar \zeta + i \zeta
\sigma^i \bar \theta,$
parametrized by Grassmann coordinates
$\zeta, \quad \bar \zeta$,
 or as a corresponding coordinate supershift
$x^{\prime i}  = x^i + i \theta\sigma^i \bar \zeta
 - i \zeta\sigma^i \bar \theta;
\qquad
\theta^{\prime}=\theta + \zeta ,\quad
{\bar\theta}^{\prime}=\bar\theta + \bar\zeta.$
\par
Assuming that coordinates $x^i$ before the supershift are the usual
c-number coordinates one sees that coordinates acquire nilpotent
Grassmann contributions after supertranslations. Therefore, there
appears a natural splitting of the space-time coordinates on the
c-number `body'-part and a nilpotent part - the so called `soul'.
The `body' subspace of superspace, or B-slice, is a submanifold
where the nilpotent part is equal to zero,
and it is a natural analogue to the real slice in complex case.
\par
Reproducing the real slice procedure of the Kerr geometry in superspace
one obtains the condition of proportionality of
the commuting spinors
$\bar\Psi(x)$  determining the PNC of the Kerr geometry
and anticommuting spinors $ \bar\theta$ and $\bar\zeta$,
As a consequence of the B-slice and superlightcone constraints
one obtains a submanifold of superspace $\theta = \theta (x),
\quad \bar \theta = \bar \theta (x).$
The initial supergauge freedom is lost now, and there appears a non-linear
realization of broken supersymmetry introduced by Volkov and Akulov
\cite{VA,WB} and considered in N=1 supergravity by Deser and Zumino 
\cite{DZ}.
It is assumed that this construction is similar to the Higgs
mechanism of the usual gauge theories, and
$\zeta ^\alpha (x), \quad \bar \zeta ^{\dot\alpha} (x) $
represent  Goldstone fermion which  can be eaten by appropriate local
supertransformation with  a corresponding redefinition of the tetrad
and the spin-3/2 field $\psi_i$.
\par
However, the complex character of
supertranslations demands to extend this scheme to N=2 supergravity.
\par
In this way the self-consistent super-Kerr-Newman solutions to broken N=2
supergravity coupled to Goldstone fermion field was obtained \cite{skn}.
 The solution describes the massless Dirac wave field propagating on
the Kerr-Newman background along the Kerr congruence.
 Besides the Kerr singular ring  solution contains an
extra axial singularity and traveling waves propagating along the ring-like
and axial singularity.
\par
The `axial' singularity represents
a half-infinite line threading the Kerr singular ring and passing to
`negative' sheet of the Kerr geometry.
The position and character of axial singularity depend on the index $n$ of
elementary excitation.
The case $n=-1/2$ is exclusive: there are two `decreasing' singularities
which are situated symmetrically at $\theta=0$ and  $\theta=\pi$.
\section{Problem of hard core}
  The obtained supergeneralization is based on the massless Goldstone
field. At present stage of investigation our knowledge regarding the origin
of the Goldstone fermion is very incompleted. Analyzing the Wess-Zumino
model of super-QED and some other schemes of spontaneously broken
supersymmetry \cite{WB}, one sees that it can leads to massless Goldstone
fermions, at least in the region of massless fields out of the BH horizons.
\par
However,  for the known parameters of spinning particles,
 the angular momentum is very high, regarding the mass parameter,  and
the black hole horizons disappear. The resulting object is "neither black
and nor hole", and the considered above disk-like `hard core' region is
naked.
Structure of this region represents a very important and extremely
complicated problem. Among the possible field models for description this
region could be mentioned the Landau-Ginzburg model, super-QED, non-abelian
gauge models, Seiberg-Witten theory, as well as the recent ideas on the
confinement caused by extra dimensions in the bulk/boundary ( AdS/CFT
correspondence) models \cite{bulk}. Apparently, this problem is very far
from resolution at present, and one of the most difficult points can be the
concordance of the field model with the rotating disk-like bag of the Kerr
geometry.
\section{Suggestions to experimental test}
 The predicted comparative big size of the disk-like bag looks as a serious
contradiction to the traditional point of view on the structureless,
point-like electron.
 However, the suggested by QED virtual photons surrounding electron in the
region of Compton size, zitterbewegung and the vacuum zero-point fluctuations,
spreading the position of electron, can be treated as some indirect evidences
for the existence of an geometrical structure in the Compton region.
At least, one can assume that region of virtual photons has a tendency to
be very ordered with formation of the Kerr congruence and the ring-like
singularity.
\par
  The modern progress in the formation of polarized beams of spinning
particles suggests the possible methods for experimental test of the main
predicted feature of the Kerr spinning particle - {\it its highly oblated
form.}
 In particular, it could be the method proposed in \cite{mus} based on
estimation of the cross section differences between
transversely and longitudinally polarized states in proton-proton
collisions. One proposes that similar experiment could be more effective
for the electron-electron collisions.
Another possible way of experimental test could be the analysis of the
diffraction of photons on the polarized electrons.
\par
Apparently, the strong influence of vacuum fluctuations will not
allow one to observe the predicted very high oblateness of electrons.
Nevertheless, one expects that an essential effect should be observed if
the Kerr source model reflects the reality.

\end{document}